\documentclass[12pt,preprint]{aastex}

\def\beq{\begin{equation}}
\def\eeq{\end{equation}}
\def\fun#1#2{\lower3.6pt\vbox{\baselineskip0pt\lineskip.9pt
  \ialign{$\mathsurround=0pt#1\hfil##\hfil$\crcr#2\crcr\sim\crcr}}}
\def\lap{\mathrel{\mathpalette\fun <}}

\shorttitle{$N$-Body Instability}
\shortauthors{Hemsendorf and Merritt}

\begin{document}

\title{Instability of the Gravitational $N$-Body Problem in the Large-$N$ 
Limit}


\author{Marc Hemsendorf and David Merritt}
\affil{Department of Physics and Astronomy, Rutgers University}

\begin{abstract}
We use a systolic $N$-body algorithm to evaluate the linear
stability of the gravitational $N$-body problem for 
$N$ up to $1.3\times 10^5$,
two orders of magnitude greater than in previous experiments.
For the first time, a clear $N$-dependence of the perturbation
growth rate is seen, $\mu_e\sim\ln N$.
The $e$-folding time for $N=10^5$ is roughly $1/20$ of a crossing time.
\end{abstract}

%\keywords{clusters: globular, peanut---bosons: bozos}

\section{Introduction}

Miller (1964) first noted the remarkable sensitivity of
the gravitational $N$-body problem to small changes in the
initial conditions.
Errors or perturbations in the coordinates or velocities of
one or more stars grow roughly exponentially, with an
$e$-folding time that is of order the crossing time.
The implication, verified in a number of subsequent studies
\citep{Lecar68,Hayli70},
is that $N$-body integrations are not reproducible over
time scales that exceed a few crossing times.
The instability is somewhat reduced when the Newtonian force
law is modified by a cutoff \citep{Standish68},
indicating that it is driven by close encounters.

Of interest is the behavior of Miller's instability in the limit of
large $N$.
It is commonly assumed that the $N$-body equations of motion
go over to the collisionless Boltzmann equation (CBE) 
as $N\rightarrow\infty$ \citep{bt87}.
This would imply, for instance, that a particle trajectory which
is integrable in a smooth potential should exhibit increasingly
regular behavior as the number of point masses used to represent
the smooth potential increases.
On the other hand, if the rate of growth of small perturbations remains 
substantial even for large $N$, there would be an important sense in which
the CBE does not correctly describe the behavior of $N$-body systems.

In fact there are indications that the growth rate of Miller's
instability remains constant or even increases with $N$
\citep{kas91,Heggie91,ghh93},
although this result is uncertain since published numerical 
experiments have been limited to $N\lap 10^3$.
Here we describe the application of a new, ``systolic''
$N$-body algorithm to this problem, which allows us
to treat systems with $N$ as large as $10^5$.
We observe for the first time a clear $N$-dependence of the
instability: the growth rate is found to increase approximately as $\ln N$.
Our methods and results are described in \S2 and \S3, and the
implications for galactic dynamics are discussed in \S4.

\section{Method}

Following Miller (1971), we integrated the coupled $N$-body and 
variational equations:
\begin{mathletters}
\begin{eqnarray}
\ddot{\bf x}_i &=& -Gm\sum_{j=1}^N 
{{\bf x}_i - {\bf x_j}\over 
|{\bf x}_i-{\bf x}_j|^3}  \label{eq_1a}
\\
\ddot{\bf X}_i &=& -Gm\sum_{j=1}^N
\left[\left({\bf X}_i - {\bf X}_j\right) - 
3{\left({\bf X}_i-{\bf X}_j\right)\cdot\left({\bf x}_i-{\bf x}_j\right)\over 
|{\bf x}_i-{\bf x}_j|^2}\left({\bf x}_i-{\bf x}_j\right)\right]
\times {1\over |{\bf x}_i-{\bf x}_j|^3}.
\label{eq_1b}
\end{eqnarray}
\end{mathletters}
Here ${\bf x}_i$ are the configuration-space coordinates of the $i$th particle
and ${\bf X}_i$ are the components of its variational vector.
The masses $m$ are assumed equal.
The variational equations represent the time development of the
infinitesimal distance between two neighboring $N$-body systems.

Equations (\ref{eq_1a}) and (\ref{eq_1b}) were integrated using the systolic $N$-body algorithm described by Dorband, Hemsendorf \& Merritt (2002).
This algorithm implements the fourth-order Hermite integration as described by Makino \& Aarseth (1992).
We adopted their formula for computing the time step of particle $i$, 
\begin{equation}
\Delta t_i = \sqrt{\eta{|{\bf a}_i(t_1)||{\bf a}_i^{(2)}(t_1)|+
|{\bf \dot a}_i(t_1)|^2\over 
|{\bf\dot a}_i(t_1)||{\bf a}_i^{(3)}(t_1)|+|{\bf a}_i^{(2)}(t_1)|^2}}.
\label{eq_tstep}
\end{equation}
Here ${\bf a}$ is the acceleration $\ddot{\bf x}$, 
superscripts denote the order of the time derivative, 
$t_1$ is the system time,
and $\eta$ is a dimensionless constant; we set $\eta=0.02$.
The same time step was used to integrate both the $N$-body and 
variational equations.
The systolic algorithm distributes the particles equally among
$p$ processors and computes forces by systematically shifting the 
particle coordinates between processors in a ring.
A single processor was used for small particle numbers while 64 
processors were used for the largest-$N$ integrations (Table \ref{tab1}).
The multi-processor integrations were carried out using the
Cray T3E at the H\"ochstleistungsrechenzentrum in Stuttgart.

\begin{figure}
\epsscale{0.9}
\plotone{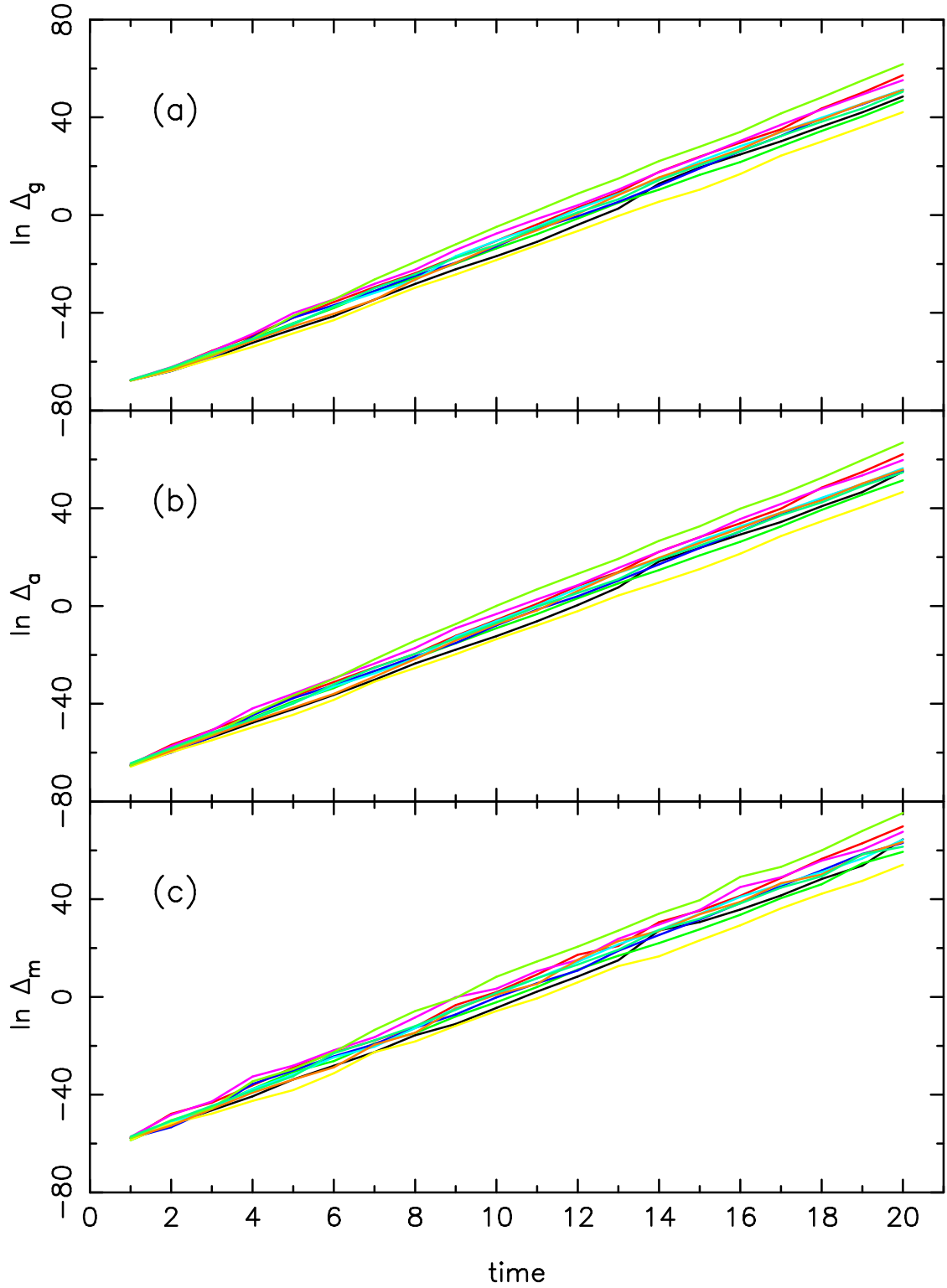}
\caption{Growth of the variation in the spatial coordinate for 10 
Plummer-model integrations with $N=32768$.
(a) $\Delta_g$, the geometric mean of the variations.
(b) $\Delta_a$, the arithmetic mean of the variations.
(c) $\Delta_m$, the maximum variation. 
\label{fig1}}
\end{figure}

Initial conditions were generated randomly from the isotropic Plummer
model, whose density and potential satisfy
\begin{equation}
\rho(r) = \frac{3 G M}{4 \pi} \; \frac{b^2}{\left(r^2 + b^2\right)^{5/2}}, 
\ \ \ \ \ \ 
\Phi(r) = - \frac{G M}{\sqrt{r^2 + b^2}}.
\end{equation}
We adopted standard $N$-body units \citep{hem86} such that $G=M=1$, $E=-1/4$,
giving a scale factor $b=3\pi/16$.
We defined the crossing time $t_{cr}$ as $R/V$, 
with $R\equiv -GM^2/2E$ and $V^2\equiv -2E/M$;
in these units, $t_{cr} = 2\sqrt{2}$.
The variational vectors ${\bf X}$ and ${\bf V}$ 
were assigned an initial amplitude of $10^{-30}$ for each particle
with randomly chosen directions.

The parameters of the integrations are listed in Table 1.
Each integration was continued until a time of 20, 
or roughly $7$ crossing times,
with the exception of the largest run, $N=131072$ which
was terminated at $t=12$.
This time is short enough that two-body relaxation should not be important
except perhaps for the smallest $N$, 
and long enough to show a clearly exponential
growth of the solutions of the variational equations.

We focussed on the amplitudes of the Cartesian components of the
variational vectors, ${\bf X}_i=(X_i, Y_i, Z_i)$,
since the variational velocities tend to exhibit spikes in their time
dependence \citep{Miller64}.
We examined three choices for the amplitude $\Delta$ of the separation:

\noindent
1. $\Delta_m$, the maximum over $i$ of the $|{\bf X}_i|$.

\noindent
2. $\Delta_a$, the arithmetic mean of the $|{\bf X}_i|$.

\noindent
3. $\Delta_g$, the geometric mean of the $|{\bf X}_i|$.

The instability growth rate $\mu_e$ was defined as
\begin{equation}
\mu_e = {\ln\Delta(t_2) - \ln\Delta(t_1)\over t_2-t_1}.
\end{equation}
Except in the case of small $N$, $N\lap 1024$, growth in the
variation was found to be nearly exponential and hence the
computed values of $\mu_e$ depended only weakly on $t_2$ and $t_1$.
We chose $t_1 = 1$ and $t_2 = 20$, except for the $N=131072$ run
for which $t_2=12$.

\begin{figure}
\epsscale{0.9}
\plotone{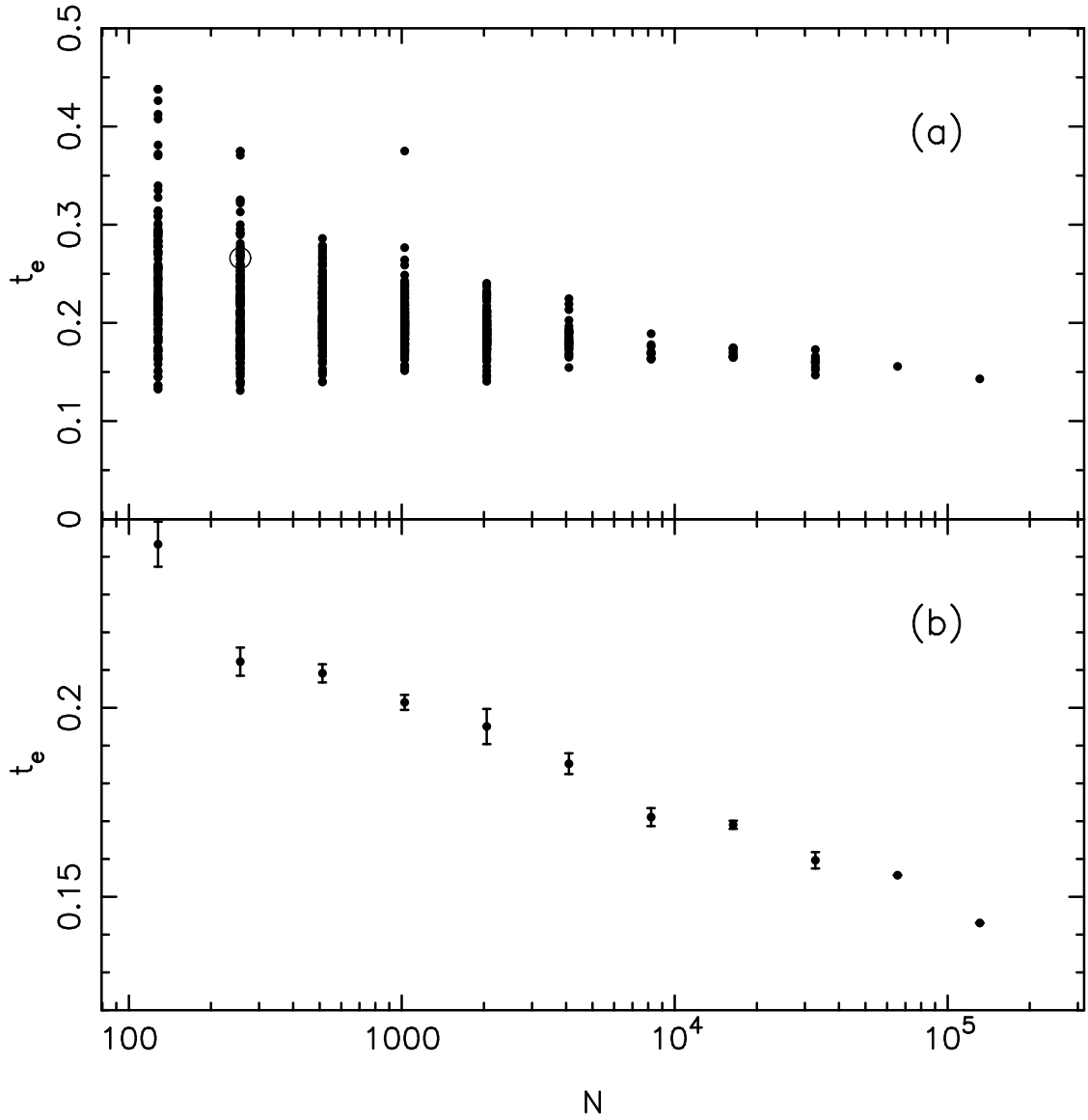}
\caption{Exponentiation times for the $N$-body integrations.
(a) $t_e$ for each of the integrations. The open circle is the
mean value of $t_e$ quoted by Goodman et al. (1993) 
from seven integrations of an $N$=256 Plummer model.
(b) Mean values of $t_e$ for each $N$.
Error bars represent the standard error of the mean.
\label{fig2}}
\end{figure}

\section{Results}

The three amplitudes $\Delta$ defined above were found to be
very similar in most of the integrations, 
as shown in Figure 1 for the 10 integrations with $N=32768$.
In what follows we adopt $\Delta \equiv \Delta_g$,
the geometric mean.

Figure 2a shows $t_e\equiv \mu_e^{-1}$ for each of the integrations.
The mean value of the $e$-folding time
and its uncertainty are plotted
in Figure 2b; the latter was defined as the standard error of
the mean, or $n^{-1/2}$ times the standard deviation, with
$n$ the number of distinct $N$-body integrations.
For the first time, a clear $N$-dependence can be seen,
in the sense that the average value of $t_e$ declines with increasing $N$:
large-$N$ systems are more unstable than small-$N$ systems.

A number of predictions have been made for
the large-$N$ dependence of the $e$-folding time.
Gurzadyan \& Savvidy (1986) estimated $t_e \sim N^{1/3}t_{cr}$
based on a geometrical approach.
This $N$ dependence is clearly inconsistent with Figure 2.
Gurzadyan \& Savvidy's approach
was criticized already by Heggie (1991) and Goodman, Heggie \& Hut (1993) 
due to its improper treatment of close encounters.
The latter authors argued that
$t_e/t_{cr}\sim 1/\ln N$ or $\sim 1/\ln(\ln N)$; 
the weaker dependence would hold only after a time long enough
that the perturbation from one star was able to propagate through
the system.

We tested these predictions against the $N$-body data.
Figure 3 shows fits of two functional forms to the mean growth rates:
\begin{mathletters}
\begin{eqnarray}
\mu_e &=& a + b\ln N, \\
\mu_e &=& c + d\ln(\ln N).
\end{eqnarray}
\end{mathletters}
Since any such relation is expected to be valid only
in the limit of large $N$, we restricted the fits to
$N\ge 1024$.
The best-fit parameters are given in Table 2.
We used a standard least-squares routine that accounts for
errors in the dependent variable ($\mu_e$); in the case of the
data pointa with $N=65536$ and 131072, for which there was only one
integration, the uncertainty in $\mu_e$ was assumed to be the
same as in the integration with $N=32768$.
These two data points were omitted when computing the values of
$\tilde\chi^2$ given in Table 2.

\begin{figure}
\epsscale{0.9}
\plotone{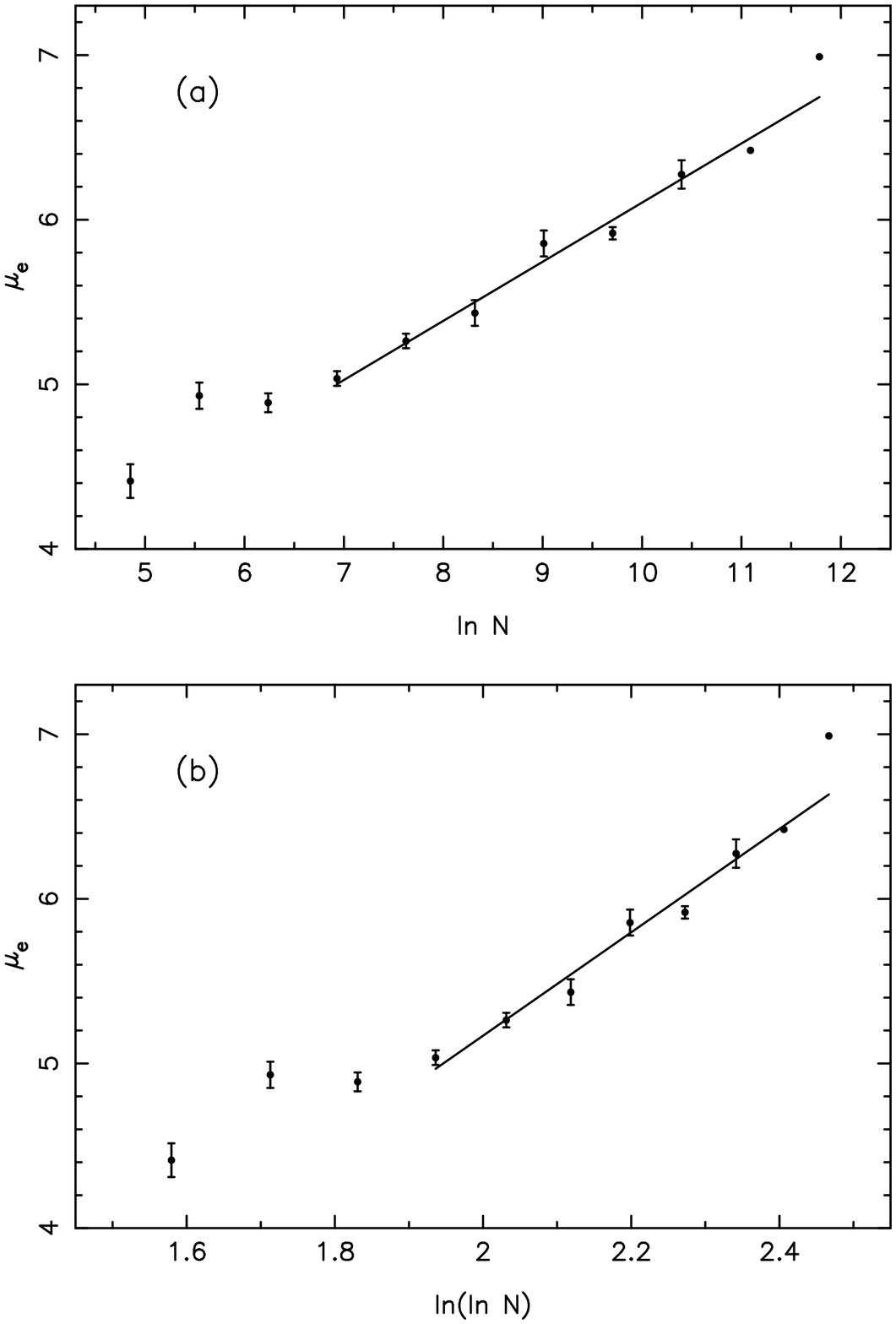}
\caption{
Fits of the $N$-body growth rates to two functional forms,
$\mu_e\propto\ln N$ and $\mu_e\propto\ln(\ln N)$.
Fitting parameters are given in Table 2.
\label{fig3}}
\end{figure}

While $\mu_e t_{cr}\propto N^0$ dependence can clearly be ruled out,
we can not distinguish between a $\ln N$ and a $\ln(\ln N)$ dependence.
Expressed in terms of $t_e$ and $t_{cr}$, the best-fit relations are
\begin{mathletters}
\begin{eqnarray}
t_e/t_{cr} &\approx& {0.99\over\ln\left(1.09\times 10^3N\right)},\\
t_e/t_{cr} &\approx& {0.112\over\ln\left(0.70\ln N\right)}.
\end{eqnarray}
\end{mathletters}
Distinguishing between these two functional forms, or other
similar ones,  would clearly be very difficult; even
for $N=10^6$, the two relations predict values of $t_e$ that
differ only by $\sim 5\%$.

\section{Discussion}

Our results confirm that the gravitational $N$-body problem
is inherently chaotic and furthermore that the degree of chaos,
as measured by the rate of divergence of nearby trajectories,
increases with increasing $N$.
We have directly established a characteristic $e$-folding time of 
$\sim t_{cr}/20$ for systems with $N\approx 10^5$;
if the weak $N$-dependence found here can be extrapolated, 
the characteristic time for systems containing $\sim10^{12}$ 
particles would be only slightly smaller, $\sim t_{cr}/30$.
Hence trajectories in stellar and galactic systems diverge on a time
scale that is generically much shorter than the crossing time.

If the collisionless Boltzmann equation (CBE) is a valid representation
of large-$N$ stellar systems,
it should be possible to show that the $N$-body trajectories
go over, in the limit of large $N$, to the orbits in the
corresponding smoothed-out potential.
The generic instability of the $N$-body problem precludes this,
since the characteristics of the CBE can not be identified
with the $N$-body orbits for times longer than $\sim t_{cr}$.
At the same time, much experience with $N$-body integrations
demonstrates that in many ways the behavior of large-$N$ systems matches  
expectations derived from the CBE (e.g. Aarseth \& Lecar 1975).

A likely resolution of this seeming paradox is that the
macroscopic, or finite-amplitude, behavior of trajectories
is not well predicted by the rate of growth of small perturbations.
For instance, orbits integrated in ``frozen'' $N$-body potentials
behave more and more like their smooth-potential counterparts
as $N$ is increased, even though their Liapunov exponents
remain large \citep{vam00,kas01}.
The initial growth of perturbations is exponential
but it saturates, at an amplitude that varies inversely
with $N$ \citep{vam00,huh01}.
Thus the CBE may be a good predictor of the macroscopic dynamics of large-$N$
systems even if it does not reproduce the small-scale chaos inherently 
associated with $N$-body dynamics.

\acknowledgments

This work was supported by NSF grant 00-71099 and by
NASA grants NAG5-6037 and NAG5-9046 to DM.
We thank S. Aarseth, J. Goodman, D. Heggie and H. Kandrup 
for useful discussions.
We are grateful to the
H\"ochstleistungsrechenzentrum in Stuttgart
for their generous allocation of computer time.

\clearpage


\begin{deluxetable}{ccccccc}
\tablecaption{Parameters of the $N$-body integrations.
\label{tab1}}
\tablewidth{0pt}
\tablehead{
\colhead{$N$} & 
\colhead{$n$}   & 
\colhead{$p$}   &
\colhead{$\langle t_e\rangle$} & 
\colhead{$\sigma_{t_e}$} &
\colhead{$\langle\mu_e\rangle$} &
\colhead{$\sigma_{\mu_e}$}}
\startdata
128   & 175 & 1  & 0.243 & 0.005 & 4.41 & 0.10 \\
256   & 175 & 1  & 0.212 & 0.004 & 4.93 & 0.08 \\
512   & 175 & 1  & 0.209 & 0.002 & 4.89 & 0.06 \\
1024  & 175 & 1  & 0.201 & 0.002 & 5.04 & 0.04 \\
2048  & 175 & 1  & 0.195 & 0.005 & 5.26 & 0.04 \\
4096  & 35 & 1  & 0.185 & 0.003 & 5.43 & 0.078 \\
8192  & 10 & 1  & 0.171 & 0.002 & 5.86 & 0.078 \\
16384 & 10 & 64 & 0.169 & 0.001 & 5.92 & 0.038 \\
32768 & 10 & 64 & 0.160 & 0.002 & 6.28 & 0.086 \\
65536 & 1  & 64 & 0.156 & ---   & 6.42 & --- \\
131072& 1  & 64 & 0.143 & ---   & 6.99 & --- \\
\enddata
\tablecomments{$N$ is the particle number;
$n$ is the number of distinct integrations; $p$ is the number of processors
used; $\langle t_e\rangle$ and $\sigma_{t_e}$ are the 
mean $e$-folding time and its error; $\langle\mu_e\rangle$ and $\sigma_{\mu_e}$ are the mean $e$-folding rate and its error.}
\end{deluxetable}

\begin{deluxetable}{cccc}
\tablecaption{Results of the linear regression fits, $Y=\alpha + bX$.
\label{tab2}}
\tablewidth{0pt}
\tablehead{
\colhead{$X$, $Y$ Variables} & 
\colhead{$\alpha$}   & 
\colhead{$\beta$}   &
\colhead{$\chi_r^2$}}
\startdata
$\ln N$, $\mu_e$                 & $2.51\pm 0.12$ & $0.36\pm 0.01$ & 2.0 \\ 
$\ln\left(\ln N\right)$, $\mu_e$ & $-1.11\pm 0.26$ & $3.14\pm 0.12$ & 3.3 
\enddata
\end{deluxetable}

\end{document}